\documentclass[letterpaper,notoc]{JHEP3}

\usepackage{graphicx,amsmath,amssymb,url}

\title{Strongly Interacting Neutrinos as the Highest Energy Cosmic Rays: A Quantitative Analysis\footnote{Talk given at the 29th Johns Hopkins Workshop on Current Problems in Particle Theory, 1-3 August 2005, Budapest, Hungary}}

\author{Markus Ahlers, \speaker{Andreas Ringwald}\\
Deutsches Elektronen-Synchrotron DESY, Notkestrasse 85, D-22607 Hamburg, Germany\\
E-mail: \email{markus.ahlers@desy.de},
        \email{andreas.ringwald@desy.de}
}
\author{Huitzu Tu\\
Department of Physics and Astronomy, University of Aarhus, Ny Munkegade, Building 1520, DK-8000 Aarhus C, Denmark\\
E-mail: \email{huitzu@phys.au.dk}
}
  
\abstract{
Scattering processes in the cosmic microwave background limit the propagation of ultra high energy charged particles in our Universe. For extragalactic proton sources resonant photopion production results in the famous Greisen-Zatsepin-Kuzmin (GZK) cutoff at about $4\times10^{10}$ GeV expected in the spectrum observed on Earth. The faint flux of ultra high energy cosmic rays of less than one event per year and cubic kilometer and the large systematic uncertainties in the energy calibration of cosmic ray showers is a challenge for cosmic ray observatories and so far the GZK cutoff has not been unambiguously confirmed. We have investigated the possibility that the primaries of super-GZK events are strongly interacting neutrinos which are not subject to the GZK cutoff. For the flux of protons and neutrinos from extragalactic optically thin sources and a flexible parameterization of the neutrino-nucleon cross section we have analyzed the cosmic ray spectra observed at AGASA and HiRes taking also into account results from horizontal events at AGASA and contained events at RICE. We find that scenarios of strongly interacting neutrinos are still compatible with the data requiring a steep increase of the inelastic neutrino-nucleon cross section by four order of magnitude within one energy decade compared to the Standard Model predictions. We also discuss the impact of the preliminary cosmic ray spectrum observed by the Pierre Auger Observatory.
}

\preprint{{\tt DESY 05-236}}

\begin{document}

\section{Introduction}

The origin and chemical composition of cosmic rays with energies above $10^9$ GeV - so called {\it ultra high energy} (UHE) cosmic rays (CRs) - is still an open question in astroparticle physics. The large-scale isotropy~\cite{Takeda:1999sg,Abbasi:2004ib} and also the shower characteristics~\cite{Bergman:2004bk} of these events give reason to believe that CRs around the {\it ankle} at about $10^{10}$ GeV are dominated by extragalactic protons. As charged particles, these protons are subject to scattering processes in the intergalactic photon gas. It has been shown~\cite{Berezinsky:2002nc} (see also Refs.~\cite{Fodor:2003bn,Berezinsky:2004fk,Ahlers:2005sn,Berezinsky:2005cq}) that the CR spectrum between $10^{9}$ GeV and $4\times10^{10}$ GeV is in a remarkably good agreement with the propagated flux from extragalactic proton sources using a simple injection spectrum with two free parameters. In this case, the ankle can be identified as an $e^+e^-$ pair production {\it dip} of protons scattering off photons of the cosmic microwave background (CMB) together with a {\it pile-up} of protons due to photo-pion production.

Soon after the discovery of the CMB in 1965 it was pointed out by Greisen~\cite{Greisen:1966jv}, Zatsepin, and Kuzmin~\cite{Zatsepin:1966jv} that resonant photopion production of protons above $4\times10^{10}$ GeV limits their propagation to about 50 Mpc.\footnote{For heavier nuclei photo-disintegration in the CMB give a comparable or even stronger attenuation above this energy.} For an extra-galactic proton source this results in a cut-off expected in the CR spectrum. So far the experimental resolution for this feature is very poor due to low statistics and large systematic errors in energy calibration (cf.\ Fig.~\ref{alldataproc}). The next generation cubic-kilometer-sized detectors, notably the Pierre Auger Observatory (PAO) will soon have the necessary exposure to resolve this issue.

A continuation of the CR spectrum  beyond the GZK cutoff to extremely high energies (EHEs) seems only consistent with a proton dominance if the sources lie in our local cosmic environment. Only very few astrophysical accelerators can achieve the necessary energies~(see e.g.\ Ref.~\cite{Torres:2004hk} for a review) and so far none of the candidate sources have been confirmed. It has been speculated that decaying superheavy particles, which could be some new form of dark matter or remnants of topological defects, could be a source of UHE CRs, but also these proposals are not fully consistent with the gamma ray spectrum at lower energies~\cite{Semikoz:2003wv}.

These missing parts of the CR {\it puzzle} has led to speculations about a different origin of EHE CRs. In particular, a flux of neutral components would not be attenuated by the CMB and could be fueled by very distant sources. In the late 60s Berezinsky and Zatsepin~\cite{Beresinsky:1969qj} proposed that cosmogenic neutrinos produced in photopion production of protons in the CMB could explain EHE events assuming a strong neutrino-nucleon interaction. The realization of such a behavior has been proposed abundantly in scenarios beyond the (perturbative) Standard Model (SM): e.g. arising through compositeness~\cite{Domokos:1986qy,Bordes:1997bt,Bordes:1997rx}, through electroweak sphalerons~\cite{Aoyama:1986ej,Ringwald:1989ee,Espinosa:1989qn,Khoze:1991mx,Ringwald:2002sw,Bezrukov:2003qm,Ringwald:2003ns,Fodor:2003bn,Han:2003ru}, through string excitations in theories with a low string and unification scale~\cite{Domokos:2000dp,Domokos:2000hm,Burgett:2004ac}, through Kaluza-Klein modes from compactified extra dimensions~\cite{Domokos:1998ry,Nussinov:1998jt,Jain:2000pu,Kachelriess:2000cb,Anchordoqui:2000uh,Kisselev:2003rz}, or through $p$-brane production in models with warped extra dimensions~\cite{Ahn:2002mj,Jain:2002kf,Anchordoqui:2002it}, respectively (for recent reviews, see Refs.~\cite{Fodor:2004tr,Anchordoqui:2005ey}).

Up to now, UHE cosmic neutrinos have been searched for in the Earth atmosphere (Fly's Eye~\cite{FLYSEYE} and AGASA~\cite{Yoshida:2001pw}), in the Greenland (FORTE~\cite{FORTE}) and Antarctic ice sheet (AMANDA \cite{AMANDA} and RICE~\cite{RICE}), in the sea/lake (notably BAIKAL~\cite{BAIKAL}), or in the regolith of the moon (GLUE~\cite{GLUE}). A large neutrino-nucleon cross section is not necessarily in conflict with the very rare UHE neutrino events claimed by these experiments, if the transition from the weak SM interaction to a strong contribution of new physics proceeds over a sufficiently short energy range. Those neutrinos interacting very strongly will then be hidden in the nucleonic CR background . In our quantitative analysis of scenarios with strong neutrino-nucleon interactions we have used the search results from AGASA and RICE.

In this paper we will report on a recent statistical analysis~\cite{Ahlers:2005zy} investigating scenarios with strong neutrino-nucleon interactions as an explanation of GZK excesses. For the flux of CRs and neutrinos from distant optically thin sources introduced in \S 2 and a model for a strong neutrino-nucleon interaction introduced in \S 3 we will show in \S 4 the results of goodness-of-fit test for existing CR data from AGASA~\cite{Takeda:2002at} and HiRes~\cite{Bird:1994wp,Abbasi:2005ni,Abbasi:2005bw,HURL} as well as preliminary data from PAO~\cite{Sommers:2005vs,PAO}. We also include the search results on neutrino-induced events from AGASA and RICE. Finally, we conclude in \S 5.

\FIGURE[t]{
\begin{minipage}[t]{\linewidth}\center
  \includegraphics[height=7cm]{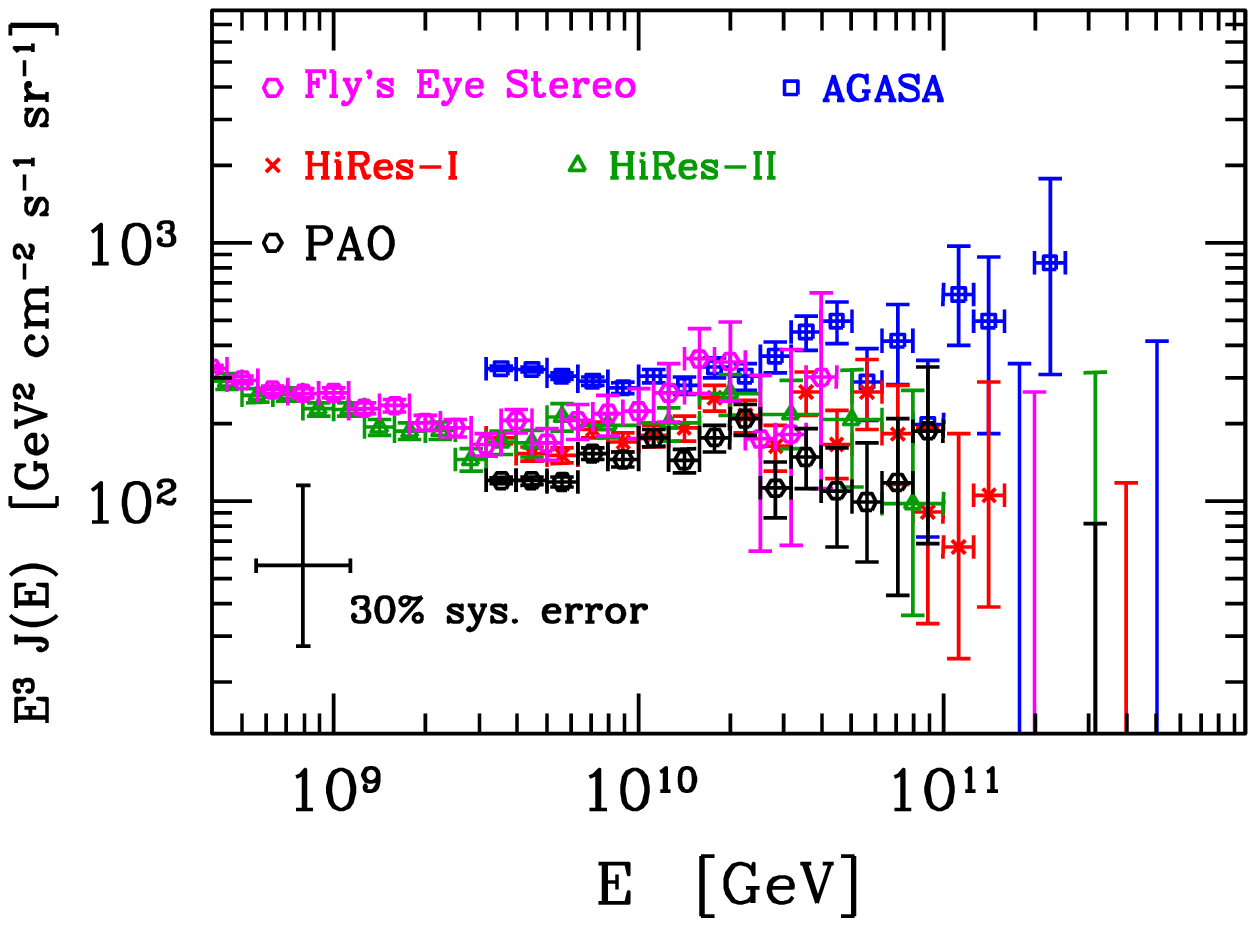}
  \caption{Cosmic ray spectrum observed by various experiments. Also shown is the size of the systematic error in energy calibration.}
\label{alldataproc}
\end{minipage}}

\section{Extragalactic Particle Production at Ultra High Energy}

The hypothesis of extragalactic proton dominance around the ankle is in very good qualitative agreement with the observed CR spectrum between $10^{9}$ GeV and $4\times10^{10}$ GeV~\cite{Berezinsky:2002nc,Fodor:2003bn,Berezinsky:2004fk,Ahlers:2005sn,Berezinsky:2005cq} as shown in Fig.~\ref{gxg}. One can distinguish two mechanisms for the production of these UHE protons. They could be produced in the decay of superheavy particles, formed in the decay of topological defects or being a part of the dark matter content in our Universe. In general, these {\it top-down} scenarios involve physics beyond the SM like Grand Unified Theories. In the more conservative {\it bottom-up} approach charged particles are magnetically confined in large astrophysical objects, such as active galactic nuclei or gamma ray bursts and accelerated through (repeated) scattering off plasma shock fronts. Photopion interactions with the photon gas inside the accelerator produces high energy neutrons, neutrinos, and gamma rays which can escape the magnetic region (see Fig.~\ref{thin_thick}). We will assume in the following that extragalactic UHE protons are produced in {\it optically thin} astrophysical accelerators, which allow for a simple determination of neutrino fluxes relative to the flux of CRs.

\FIGURE[t]{
\begin{minipage}[t]{\linewidth}\center
  \includegraphics[height=6cm]{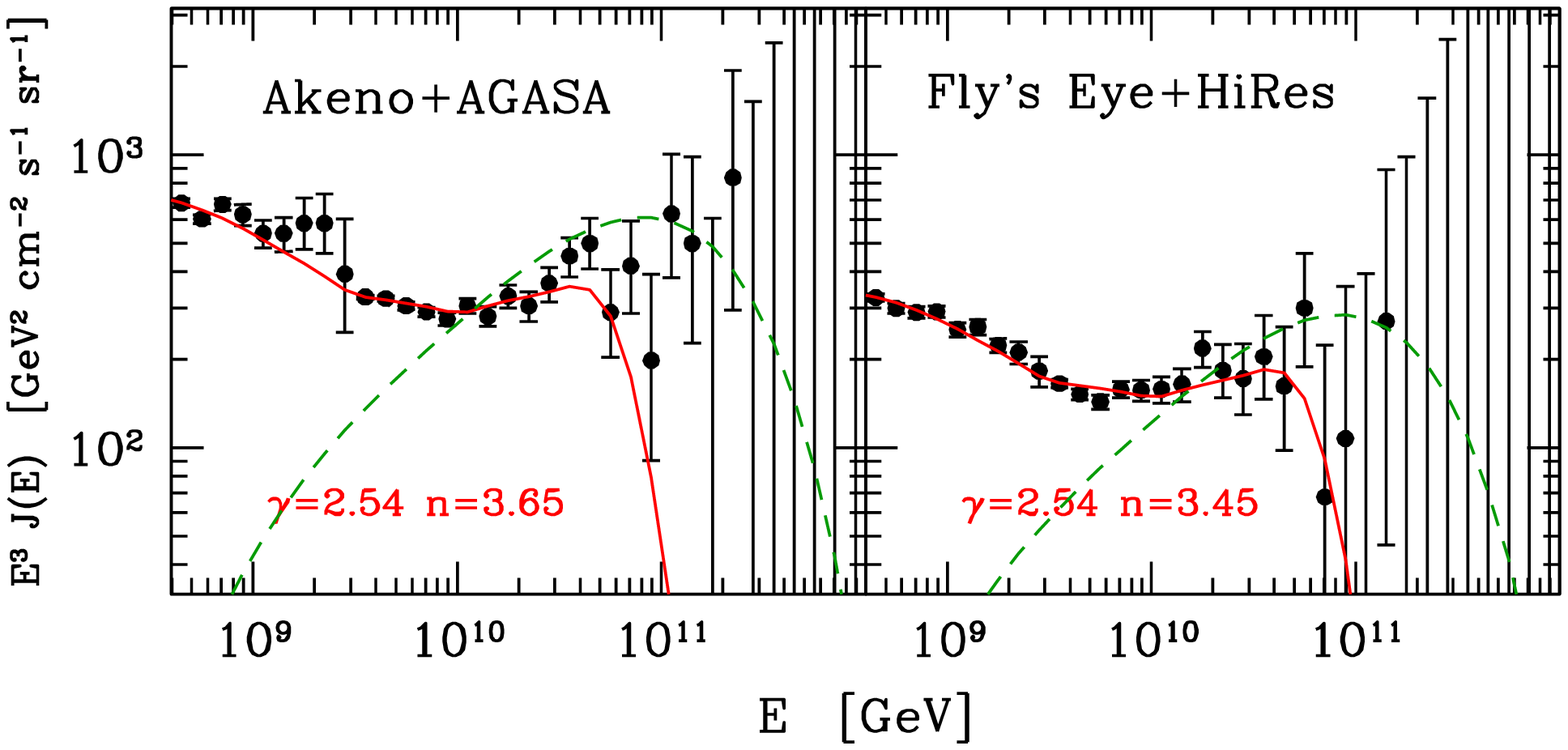}
  \caption{The flux of extragalactic protons (solid line) from Ref.~\cite{Ahlers:2005sn} using a power-like injection spectrum with red-shift evolution of the source luminosity $\propto(1+z)^nE^{-\gamma}$. Also shown are the fluxes of cosmogenic neutrinos (dashed line) summed over all flavors.}\label{gxg}
\end{minipage}
}

\FIGURE[b]{
\begin{minipage}[t]{\linewidth}
\begin{minipage}[t]{0.48\linewidth}\center
  \includegraphics[height=6cm,clip=true]{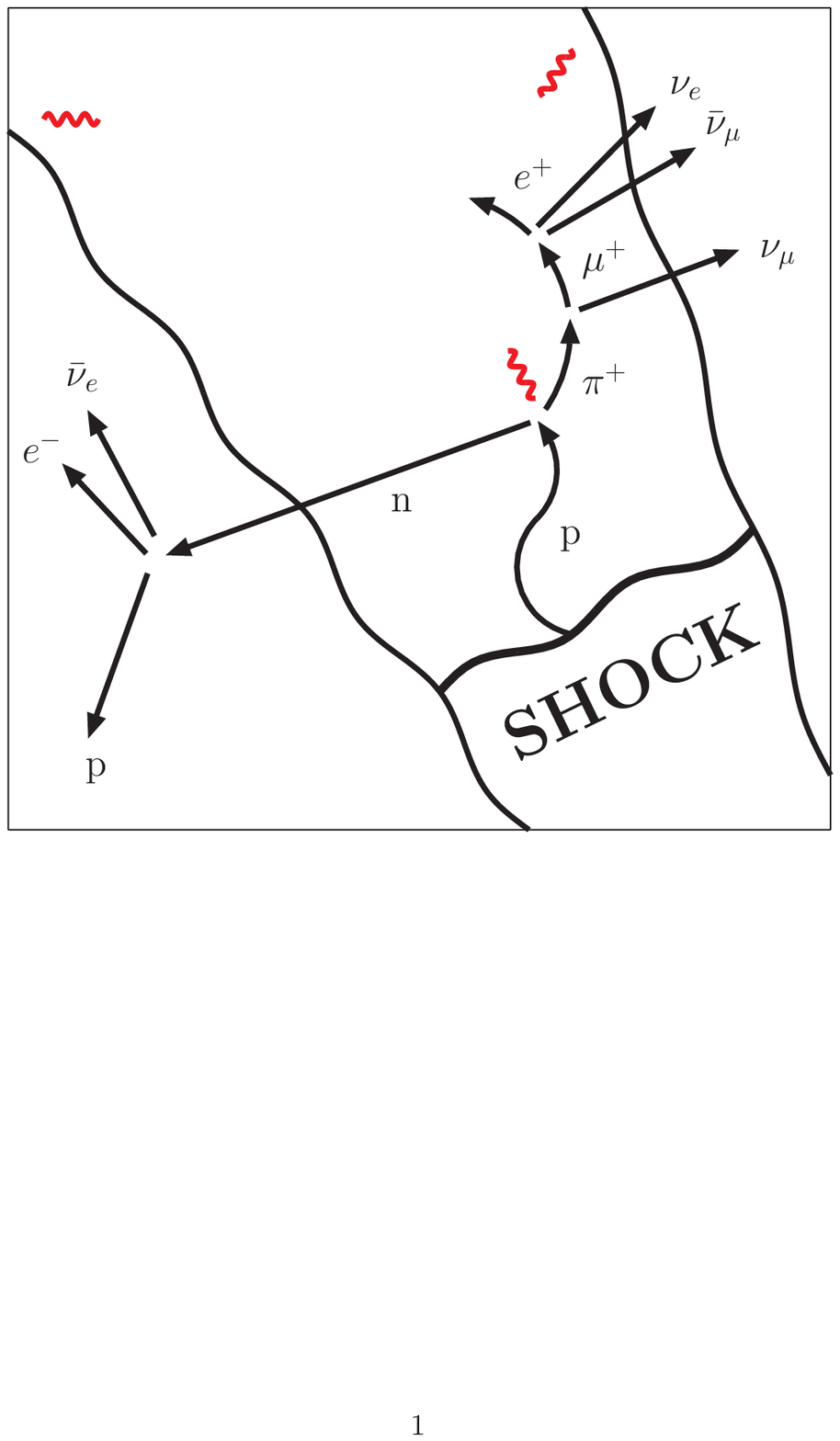}
\end{minipage}
\hfill
\begin{minipage}[t]{0.48\linewidth}\center
  \includegraphics[height=6cm,clip=true]{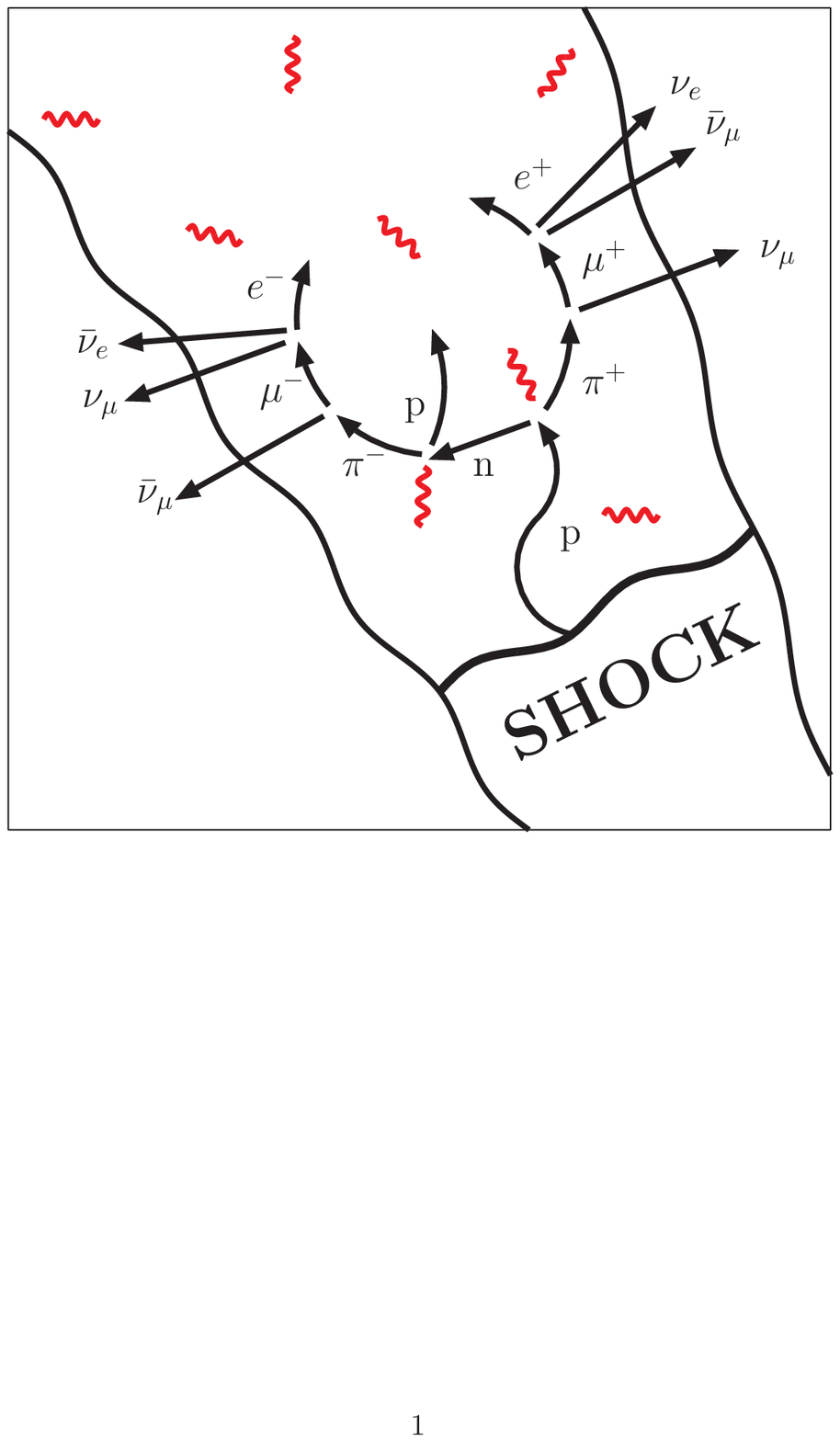}
\end{minipage} \caption{{\bf Left panel:} Sketch of an {\it optically thin} CR source. Each UHE proton is accompanied by three UHE neutrinos. {\bf Right panel:} In {\it optically thick} CR sources neutrons may scatter of the photon gas before escaping the source. This produces a larger neutrino to CR ratio.}\label{thin_thick}
\end{minipage}
}

\FIGURE[t]{
\begin{minipage}[t]{\linewidth}\center
  \includegraphics[height=8cm]{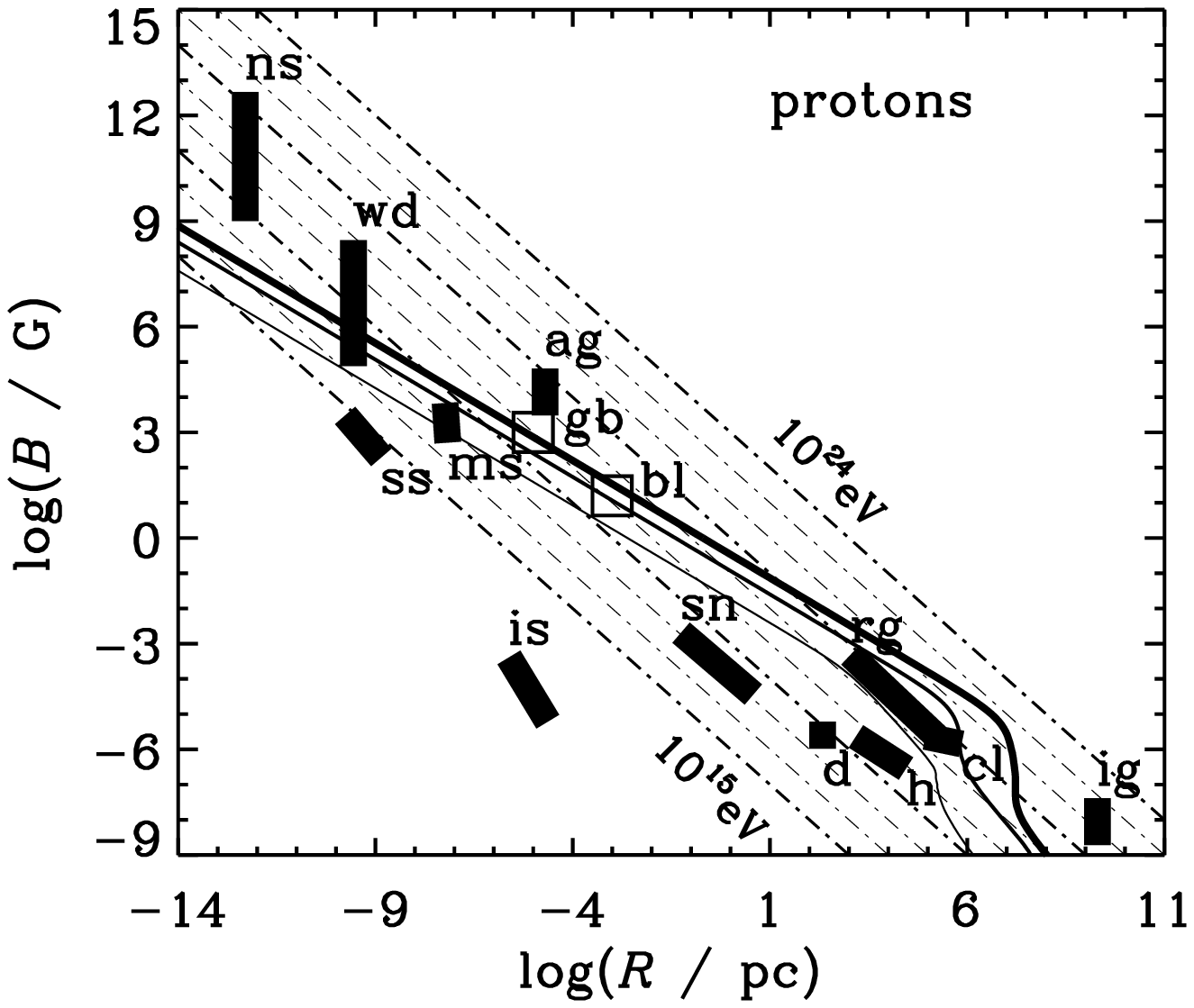}
  \caption{{\it Hillas diagram} (from Ref.~\cite{Protheroe:2004rt}) for objects with characteristic radius $R$ and magentic field $B$.  The dashed-dotted lines correspond to different proton energies from $10^{6}$ GeV to $10^{15}$ GeV. The solid lines correspond to limits of the maximal proton energy including energy losses and interactions for the models discussed in Ref.~\cite{Protheroe:2004rt}. The sources are: neutron stars (ns), white dwarfs (wd), sunspots (ss), magnetic stars (ms), active galactic nuclei (ag), interstellar space (is), supernova remnants (sn), galactic disk (d), halo (h), radio galaxy lobes (rg), clusters of galaxies (cl) and intergalactic medium (ig). Also shown are jet-frame parameters for blazars (bl) and gamma ray bursts (gb).}\label{hillas}
\end{minipage}
}

\subsection{Extragalactic Protons and Cosmogenic Neutrinos}

For an efficient acceleration of a charged particle by repeated scattering processes within an astrophysical source it is necessary that its gyroradius is smaller than the size of the accelerator. This simple geometric argument is conveniently depicted as the {\it ``Hillas diagram''} shown in Fig.~\ref{hillas}. In the case of non-cyclic one-shot acceleration scenarios a similar expression for the maximal energy holds ~\cite{Torres:2004hk}. From Fig.~\ref{hillas} one can see that only very few objects are able to produce UHE protons and even less of them EHE protons as candidates for super-GZK events. In our statistical analysis we have assumed that UHE protons are injected by spatially homogeneous and isotropic distributed astrophysical accelerators with red-shift dependent luminosity according to
\begin{gather}\label{fluxp}
  \mathcal{L}_{\rm CR}({z,E_{\rm CR}}) \propto{(1+z)^n}\,{E_{\rm CR}^{-\gamma}\,e^{-\frac{E_{\rm CR}}{E_\mathrm{max}}}}\\
 z_{\rm min}<z<z_{\rm max}\nonumber
\end{gather}
We exclude nearby (redshift $z_\mathrm{min}$) and early ($z_\mathrm{max}$) sources and fix these parameters in the following at $z_\mathrm{min} = 0.012$ (corresponding to $r_\mathrm{min} \approx 50\, \mathrm{Mpc}$) and $z_\mathrm{max} = 2$.  The maximal injection energy $E_\mathrm{max}$ is fixed at $10^{12}$~GeV in our analysis which is compatible with the highest energy proton sources shown in the Hillas diagram Fig.~\ref{hillas}.

The flux of UHE protons from distant sources is subject to energy redshift, $e^+e^-$ pair production, and photopion-production in the CMB (see e.g.~\cite{Engel:2001hd,Fodor:2003ph,Semikoz:2003wv}) which can be taken into account by means of propagation functions~\cite{Fodor:2000yi,Fodor:2003bn}. The resonantly produced pions, which are also the reason for the GZK cut-off, decay into electron and muon neutrinos via the reaction chain $p + \gamma_{\rm CMB} \rightarrow N + \pi^+ \rightarrow \mu^+\nu_\mu \ldots  \rightarrow\nu_\mu\bar \nu_\mu \nu_e\, e^+\ldots$ and provide a guaranteed source of {\it cosmogenic} UHE neutrinos observed at Earth (see Fig.~\ref{gxg}). The initial ratio of neutrinos $\nu_e:\nu_\mu:\nu_\tau$ as 1:2:0 is then redistributed as 1:1:1 by oscillation effects. 

\subsection{Neutrinos from Optically Thin Sources}

The inelastic scattering of the beam protons off the ambient photon gas in the source will also produce photopions which provide an additional source of UHE neutrinos (see Fig.~\ref{thin_thick}). In general, the relative production of other (neutral) particles in the source depend on details such as the densities of the target photons and the ambient gas~\cite{Mannheim:1998wp}. In the case of {\it optically thin} proton sources we can estimate the flux of UHE neutrinos as
\begin{gather}\label{fluxnu}
  \mathcal{L}_\nu(z,E_\nu)\approx 3
  \left\langle\frac{E_{\rm CR}}{E_\nu}\right\rangle\mathcal{L}_{\rm CR}
  \left(z,\left\langle \frac{E_{\rm CR}}{E_\nu}\right\rangle E_\nu\right).
\end{gather}

\FIGURE[t]{
\begin{minipage}[t]{\linewidth}\center
  \includegraphics[width=0.6\linewidth,clip=true]{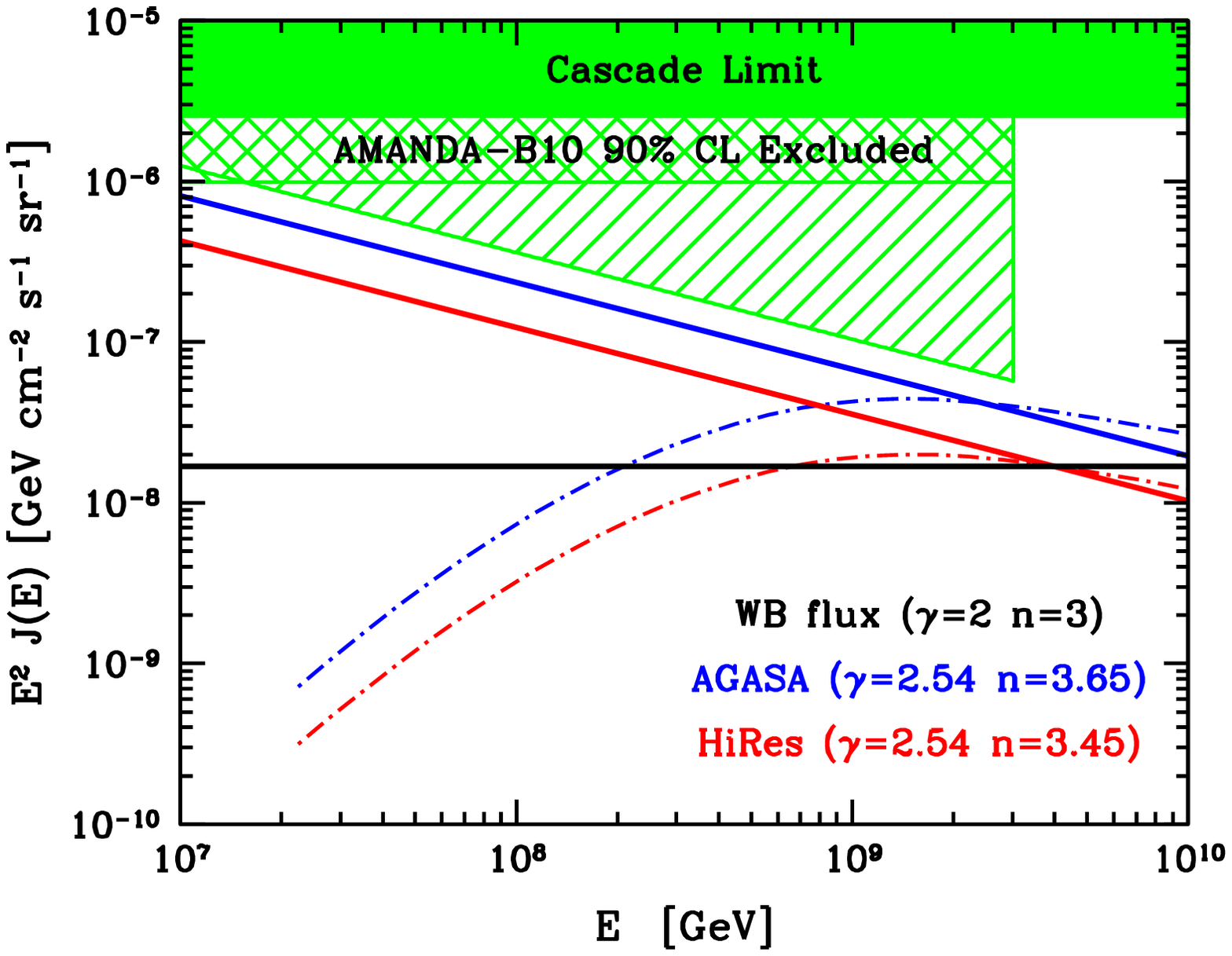}
  \caption{The neutrinos from optically thin sources (tilted solid lines) are close to be measured by neutrino telescopes such as AMANDA/IceCube (from Ref.~\cite{Ahlers:2005sn}). The fluxes are normalized to the AGASA and HiRes cosmic ray spectrum, respectively. Also shown are the Waxman-Bahcall flux~\cite{Waxman:1998yy} (horizontal solid line) assuming a CR injection as $E^{-2}$ and the fluxes of cosmogenic neutrinos (dashed-dotted lines). }
\label{thinflux}
\end{minipage}
}

For the mean energy ratio we have used $\langle E_{\rm CR}/E_\nu\rangle\approx0.07$, suitable for resonant photoproduction at the energies in question (see Ref.~\cite{Ahlers:2005sn}). This flux of neutrinos is directly associated with the injected flux of CRs and serves as a minimal contribution from the source. In optically thicker sources neutrons may undergo photohadronic interactions before escaping the confinement region as it is sketched in the right panel of Fig.~\ref{thin_thick}. This will decrease the emissivity of neutrons compared to that of neutrinos.  Depending on the ambient gas, $pp$ interactions may also dominate over photohadronic processes in the source and produce additional neutrinos. As it was discussed in Ref.~\cite{Ahlers:2005sn} this flux of high energy neutrinos from optically thin sources is almost in reach of the AMANDA-II detector for neutrino energies of the order $10^7$~GeV and should soon be observable by its successor experiment IceCube (see Fig.~\ref{thinflux}). This observation is crucial for our assumption, that extragalactic protons start to dominate in the CR spectrum already at a low crossover below the ankle. If even cosmogenic neutrinos are not detected in future experiments, also the hypothesis of extragalactic proton dominance above the ankle has to be reconsidered.

\section{Strongly Interacting Cosmic Neutrinos}

The SM interaction length of neutrinos exceeds the vertical column depth by several orders of magnitude. For a reasonable contribution of extragalactic neutrinos in vertical CRs one has to assume a strong enhancement of the neutrino-nucleon interaction. However, the transition from a weak to a strong interaction has to be sufficiently fast in order to reproduce the low neutrino rates or negative search results from experiments such as AGASA~\cite{Yoshida:2001pw}, AMANDA~\cite{AMANDA}, ANITA~\cite{ANITA}, BAIKAL~\cite{BAIKAL}, Fly's Eye~\cite{FLYSEYE}, FORTE~\cite{FORTE}, GLUE~\cite{GLUE}, and RICE~\cite{RICE}. Further ambitious projects
include IceCube~\cite{ICECUBE} and PAO~\cite{PAO_hor}, possibly followed by EUSO~\cite{EUSO}, OWL~\cite{OWL}, or SalSA~\cite{SALSA}.

\FIGURE[b]{
\begin{minipage}[t]{\linewidth}
\begin{minipage}[t]{0.48\linewidth}\center
  \includegraphics[width=\linewidth]{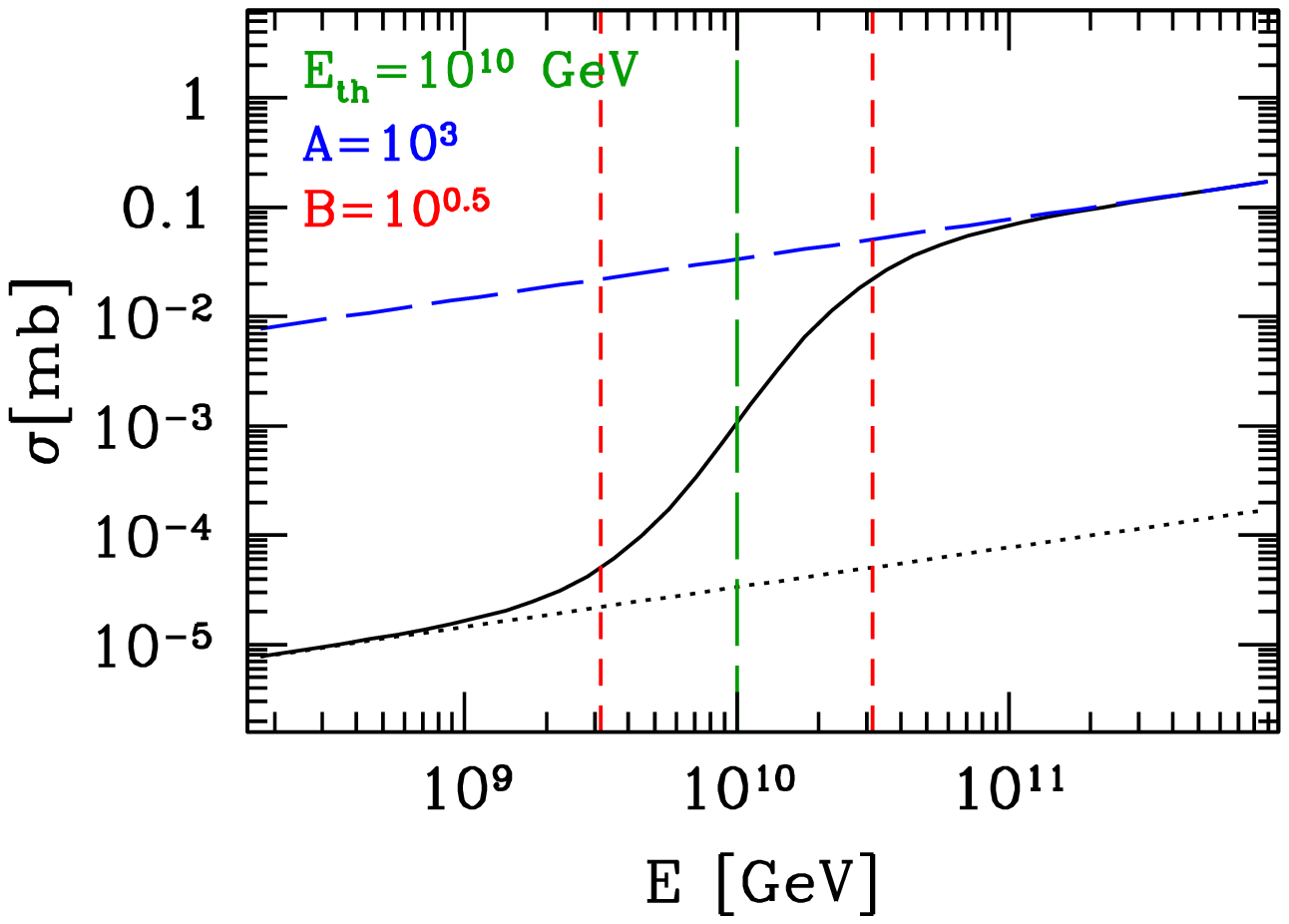}
\end{minipage}
\hfill
\begin{minipage}[t]{0.48\linewidth}\center
  \includegraphics[width=\linewidth]{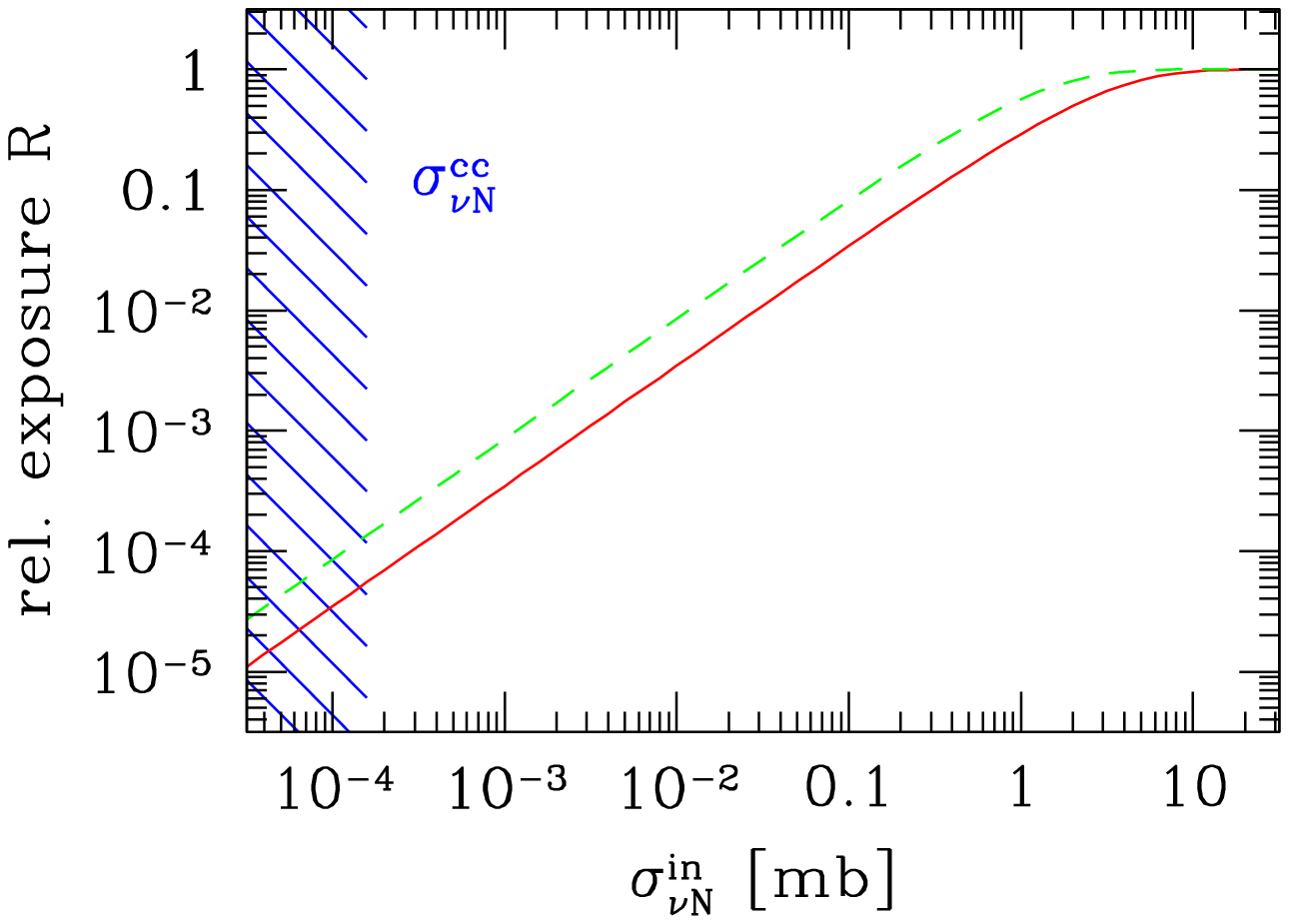}
\end{minipage}
\caption{{\bf Left panel:} An example of the strong neutrino-nucleon interaction given in Eq.~(\protect\ref{cs}). {\bf Right panel:} Relative exposure $\mathcal{R}$ (Eq.~(\protect\ref{rel_exposure})) to strongly interacting neutrinos as a function of the neutrino-nucleon inelastic cross section $\sigma^{\rm in}_{\nu N}$, of AGASA (solid) and HiRes (dashed), respectively).  The {hatched} region shows the predicted contribution from SM charged current interaction~\cite{Gandhi:1998ri,Kwiecinski:1998yf}.}
\label{cs_example_stoR}
\end{minipage}
}

In the introduction we have already given examples of scenarios beyond the (perturbative) SM which predicts a strong neutrino-nucleon interaction. In the following we will use a flexible parameterization of a strong neutrino-nucleon inelastic cross section ($\sigma_{\nu N}^{\rm new}$) focusing on three characteristic parameters: (i) the energy scale $E_{\rm th}$ of the new underlying physics, (ii) the amplification $\mathcal A$ compared to the SM predictions, and (iii) the width $\mathcal B$ of the transition between weak and strong interaction. A mathematical convenient parameterization illustrated in Fig.~\ref{cs_example_stoR} is given by

\begin{equation}\label{cs}
  \log_{10}\left(\frac{\sigma_{\nu N}^{\rm new}}{\mathcal{A}\,\sigma_{\nu N}^{\rm SM}}\right) =
  \frac{1}{2}\left[1+\tanh\left(\log_{\cal
            B}\frac{E_\nu}{E_{\rm th}}\right)\right].
\end{equation}
In general, experiments distinguish different CR primaries by their characteristic shower development in matter. For the sake of simplicity we will assume that the characteristics of the showers induced by strongly interacting neutrinos are indistinguishable from those induced by protons.  In particular, we assume for both primaries (i) a complete conversion of the incident energy into the shower, and (ii) equal detection efficiencies at the highest energies.

Under these assumptions the number of detected events can be expressed as
\begin{equation}
\label{rel_exposure}
  N_\mathrm{obs} = \int d\,E\,\mathcal{E}(E)\left( J_p (E)+
  \mathcal{R}(\sigma^\mathrm{new}_{\nu N} (E))\, J_\nu(E)\right)\, ,
\end{equation}
where we define $\mathcal{R}(\sigma_{\nu N}^\mathrm{new})$ as the relative experimental exposure to strongly interacting neutrinos compared to the exposure $\mathcal{E}(E)$ to protons. In general, the value of $\mathcal{R}(\sigma^\mathrm{new}_{\nu N}(E))$ is determined by the search criterion on the zenith angle $\theta$ and the (observed) atmospheric depth adopted by each experiments. Figure \ref{cs_example_stoR} gives the relative exposure for vertical showers at AGASA and HiRes (see Ref.~\cite{Ahlers:2005zy} for details).

\section{Quantitative Analysis}

For the vertical CR spectrum of AGASA, HiRes-I/II, and the Fly's Eye Stereo data together with the search results of horizontal events at AGASA and contained events at RICE we show in the left panel of Fig.~\ref{csbound} the 90\%, 95\%, and 99\% confidence level (CL) of the inelastic cross section (Eq.~\ref{cs}) from a goodness-of-fit test using the fluxes from optically thin sources (Eq.~\ref{fluxp} and \ref{fluxnu}). The details of the statistical analysis and the approximations involved can be found in Ref.~\cite{Ahlers:2005zy}. 
The goodness-of-fit test of our model using the combined data requires, to the 90\% CL, a steep increase by an amplification factor of $\mathcal{A}>10^4$ over a tiny energy interval $\mathcal{B}<10^{0.5}$ at about $10^{11}$~GeV.  Within this particular scenario of strongly interacting neutrinos, the SM neutrino-nucleon inelastic cross section corresponding to $\mathcal{A}=1$ is not favored by the data.

\FIGURE[t]{
\begin{minipage}[t]{\linewidth}
\begin{minipage}[t]{0.48\linewidth}\center
  \includegraphics[width=\linewidth]{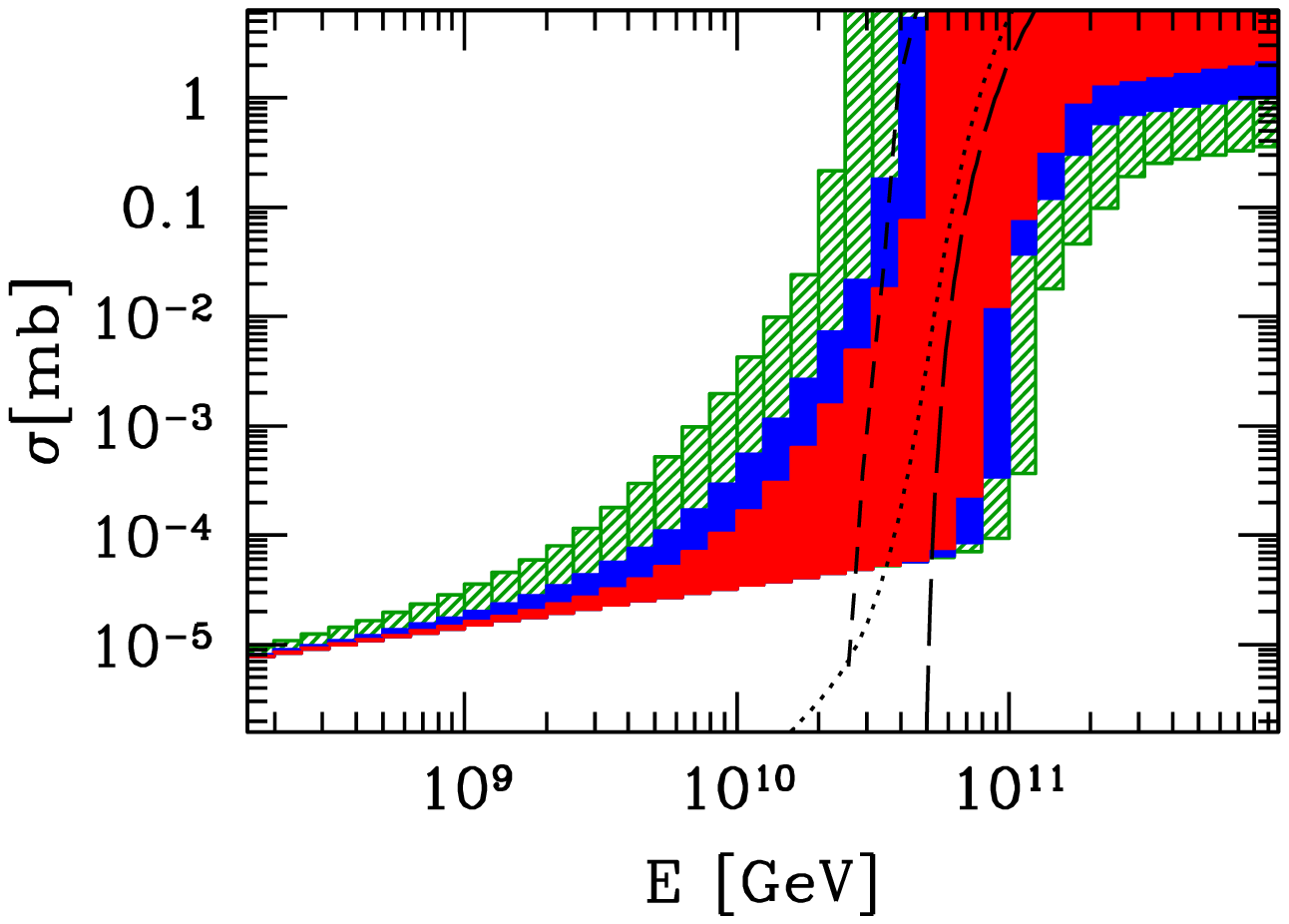}
\end{minipage}
\hfill
\begin{minipage}[t]{0.48\linewidth}\center
  \includegraphics[width=\linewidth]{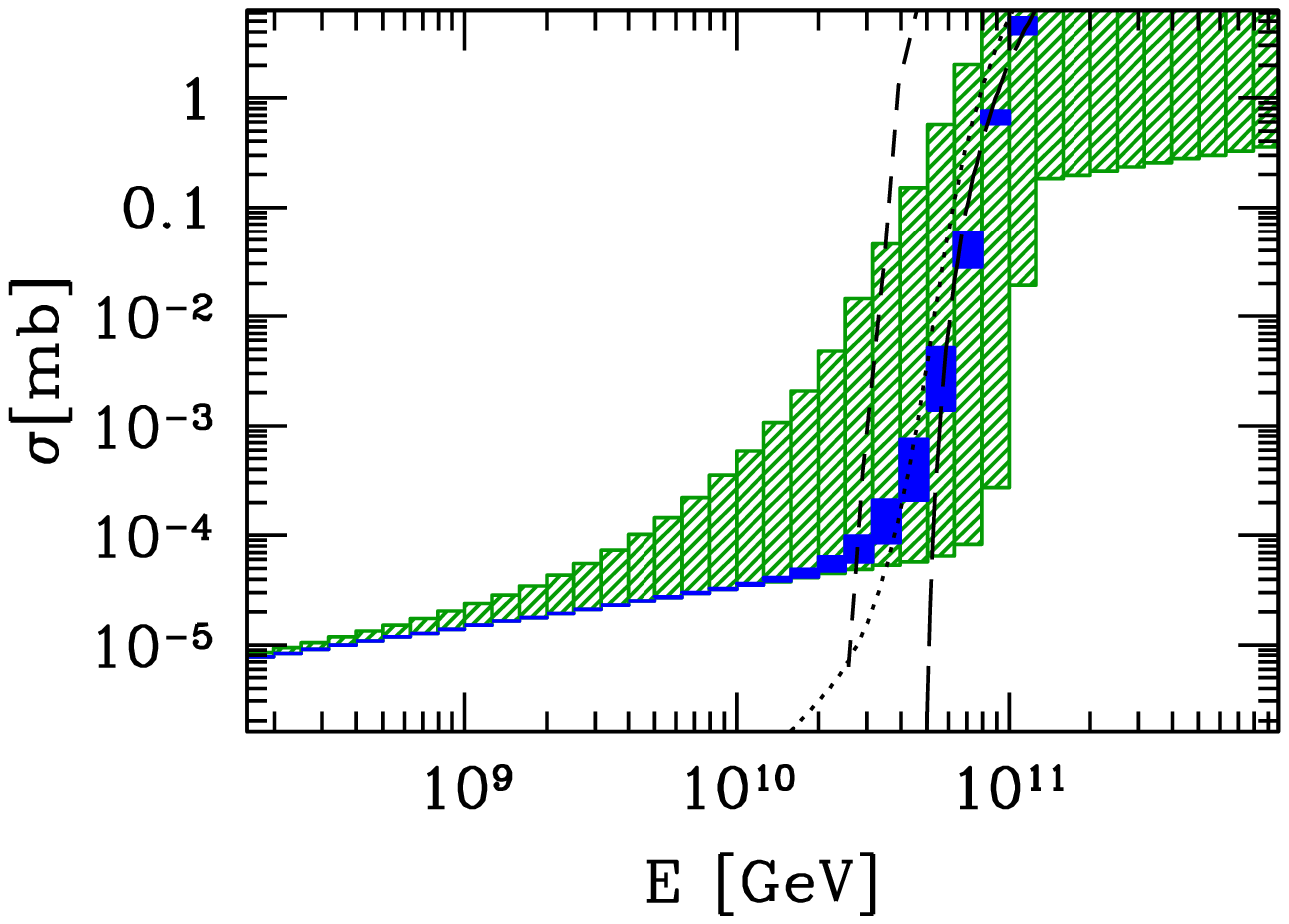}
\end{minipage}
\caption{The range of the cross section within the 99\%, 95\% and 90\% CL. The lines are theoretical predictions of an enhancement of the neutrino-nucleon cross-section by electroweak sphalerons~\cite{Han:2003ru} (short-dashed), $p$-branes~\cite{Anchordoqui:2002it} (long-dashed) and string excitations~\cite{Burgett:2004ac} (dotted). {\bf Left panel}: Fit to the AGASA and HiRes CR data. {\bf Right panel}: Fit to the AGASA, HiRes, and Pierre Auger CR data.}
\label{csbound}
\end{minipage}
}

As an illustration, we have considered three models of a rapidly increasing neutrino-nucleon cross section based on electroweak sphalerons~\cite{Han:2003ru}, $p$-branes~\cite{Anchordoqui:2002it} and string excitations~\cite{Burgett:2004ac}. (i) {\it Electroweak sphalerons:} We have used the neutrino-nucleon cross section induced by electroweak instantons shown in Ref.~\cite{Han:2003ru}, based on the neutrino-parton cross section from Ref.~\cite{Ringwald:2003ns}, the latter exploiting numerical results from Ref.~\cite{Bezrukov:2003qm}. A fit of the injection spectrum and the red-shift evolution of the source luminosity using this cross section reproduces a 98\% CL of this model, which is in very good agreement with Fig.~\ref{csbound}. (ii) {\it $p$-branes:} We have calculated $\sigma(\nu N \rightarrow brane)$ given in Ref.~\cite{Anchordoqui:2002it} as Eqs.~(9) and (10) for $m=6$ extra spatial dimensions, a fundamental scale of gravity $M_D = 300$~TeV and a ratio $L/L_* = 0.005$ of small to large compactification radii. We let all partons interact universally with the neutrino. Our fit of $\gamma$ and $n$ gives a 83\% CL. (iii) {\it String excitations:} For the neutrino-quark cross section given in Ref.~\cite{Burgett:2004ac} with a string scale $M_* = 70$~TeV and for the set of parameters $N_0 = C = 16$, characterizing the width and the absolute normalization, respectively (cf.\ their Fig.~2), we derived the neutrino-nucleon cross section.  Our fit gives 84\% CL for this cross section, again in very good agreement with Fig.~\ref{csbound}.

In a separate fit we have also included the preliminary CR data reported from PAO~\cite{PAO}. The result is shown in the right panel of Fig.~\ref{csbound}. The rate of horizontal events from PAO will also have a great impact on the analysis, but preliminary results are not yet published. The combined vertical spectrum of AGASA, HiRes, and PAO is still compatible at the 95\% CL with our scenario as one can see in the right panel of Fig.~\ref{csbound}.

\section{Conclusions}

Our analysis shows that ultra high energy cosmic rays measured at AGASA and HiRes might be composed of extragalactic protons and strongly interacting neutrinos. The enhancement of the neutrino-nucleon cross section has to be rapid in order to agree with the experimental results from horizontal events at AGASA and contained events at RICE. We derived allowed regions of a flexible parameterization of the cross section using a goodness-of-fit test and showed the compatibility of our results with theoretical predictions from sphalerons, $p$-branes and string excitations. The good agreement between these fits with our simple parameterization indicate that out results (in particular Fig.~\ref{csbound}) can be used as a {\it quick test} for strongly interacting neutrino-nucleon cross sections.

Our results are based on the assumption that ultra high energy protons and neutrinos originate at optically thin sources. One should keep in mind that a more general injection spectrum of cosmic rays or optically thicker sources would certainly alter the prediction of the neutrino fluxes and accordingly our results for the range of neutrino-nucleon interactions. As a rule of thumb cosmic accelerators always produce cosmic rays, neutrinos and gamma rays with comparable luminosities. Also the top-down production of cosmic rays is always accompanied with high energy fluxes of other particles. Hence, the search for the origin of the highest energy cosmic rays involves essentially a multi-messenger analysis. It is in the focus of future neutrino experiments like ANITA, IceCube or the Pierre Auger Observatory (horizontal events) and gamma ray observatories like H.E.S.S.~\cite{HESS} and MAGIC~\cite{MAGIC} to resolve some of these model ambiguities in the analysis.

Notably the Pierre Auger Observatory combining the experimental techniques of AGASA and HiRes as a hybrid detector will have a great impact on scenarios with strongly interacting neutrinos. With a better energy resolution and a much higher statistics it will certainly help to clarify our picture on ultra high energy cosmic rays.


\providecommand{\href}[2]{#2}\begingroup\raggedright
\endgroup
\end{document}